\documentclass{ws-ijmpd}
\begin{document}

\def\cdash{$^{\raisebox{-0.5pt}{\hbox{--}}}$}       

%

\def\nocropmarks{\vskip5pt\phantom{cropmarks}}

\let\trimmarks\nocropmarks      


    
\def\beq{\begin{eqnarray}}
\def\eeq{\end{eqnarray}}

\def\s{\mbox{\boldmath$\displaystyle\mathbf{\sigma}$}}

\def\be{\mbox{\boldmath$\displaystyle\mathbf{\eta}$}}

\def\bp{\mbox{\boldmath$\displaystyle\mathbf{\pi}$}}

\def\J{\mbox{\boldmath$\displaystyle\mathbf{J}$}}

\def\K{\mbox{\boldmath$\displaystyle\mathbf{K}$}}

\def\P{\mbox{\boldmath$\displaystyle\mathbf{P}$}}

\def\p{\mbox{\boldmath$\displaystyle\mathbf{p}$}}

\def\hp{\mbox{\boldmath$\displaystyle\mathbf{\widehat{\p}}$}}

\def\x{\mbox{\boldmath$\displaystyle\mathbf{x}$}}

\def\0{\mbox{\boldmath$\displaystyle\mathbf{0}$}}

\def\bv{\mbox{\boldmath$\displaystyle\mathbf{\varphi}$}}

\def\bx{\mbox{\boldmath$\displaystyle\mathbf{\xi}$}}

\def\bs{\mbox{\boldmath$\displaystyle\mathbf{\sigma}$}}

\def\bc{\mbox{\boldmath$\displaystyle\mathbf{\chi}$}}

\def\hbv{\mbox{\boldmath$\displaystyle\mathbf{\widehat\varphi}$}}

\def\hbxi{\mbox{\boldmath$\displaystyle\mathbf{\widehat\xi}$}}

\def\bn{\mbox{\boldmath$\displaystyle\mathbf{\nabla}$}}

\def\bl{\mbox{\boldmath$\displaystyle\mathbf{\lambda}$}}

\def\br{\mbox{\boldmath$\displaystyle\mathbf{\rho}$}}

\def\openone{\mbox{\boldmath$\displaystyle\mathbf{I}$}}

\def\1{1}

\def\ar{\stackrel{\hspace{0.04truecm}grav. }{\mbox{$\longrightarrow$}}}


\markboth{D. V. Ahluwalia-Khalilova}{Special relativity with two 
invariant scales: Motivation, Fermions, Bosons, Locality, and Critique}

%
\catchline{}{}{}
%

\title{SPECIAL RELATIVITY WITH TWO INVARIANT SCALES:\\
MOTIVATION, FERMIONS, BOSONS, LOCALITY, AND CRITIQUE\footnote{The work presented here is an extended version of the
informal lectures given by the author in May 2001 in Rome. }}


\author{\footnotesize D. V. AHLUWALIA-KHALILOVA\footnote{E-mail: 
d.v.ahluwalia-khalilova@heritage.reduaz.mx}}

\address{Center for Studies of Physical, Mathematical, and
Biological Structure of Universe\\
Department of Mathematics, Ap. Postal C-600\\
University of Zacatecas (UAZ), Zacatecas, ZAC 98062, Mexico}
\maketitle
\begin{abstract}
We present a Master equation for description of fermions and bosons
for special relativities with two invariant 
scales, \texttt{SR2}, ($c$ and $\lambda_P$). 
We introduce canonically-conjugate  
variables $\left(\chi^0, \bc\right)$ to 
$(\epsilon,\pi)$ of Judes-Visser. 
Together, they bring in a formal element of linearity and 
locality in an otherwise non-linear and non-local theory.
Special relativities with two invariant scales provide \textit{all} 
corrections, say, to the standard model of the high energy physics, 
in terms of \textit{one} 
fundamental constant, $\lambda_P$. It is emphasized that 
spacetime of special relativities with two invariant scales 
carries an intrinsic quantum-gravitational character. 
In an addenda, we also comment on the physical importance of a phase factor
that the whole literature on the subject has missed and present a 
brief critique of \texttt{SR2}.
In addition, we remark that the most natural and  physically viable 
\texttt{SR2}
shall require momentum-space and spacetime to be
non-commutative with the non-commutativity 
determined by the spin content and C, P, and T properties of the
examined representation space.  
Therefore,
in a physically successful \texttt{SR2},
the notion of spacetime is expected to be deeply 
intertwined with specific properties of the test particle.
\end{abstract}



\section{Motivation for \texttt{SR2}}

There is a growing theoretical evidence that gravitational
and quantum frameworks carry some elements of incompatibilities. 
The question is how deep are the indicated changes, and 
what precise form they may take. One hint comes from 
the observation that incorporating gravitational effects in 
quantum measurement of spacetime events leads to a Planck-scale
saturation. In the framework of Kempf, Mangano,  Mann, and the 
present author,\cite{Kempf:1994su,Ahluwalia:2000iw} 
the gravitationally-induced modification to the 
de Broglie (dB) wave-particle duality 
takes the form\cite{Ahluwalia:2000iw} 
\beq
\quad \lambda_{dB}=\frac{h}{p}
\quad \ar\quad
\lambda = \frac{\overline{\lambda}_P}
{\tan^{-1}(\overline{\lambda}_P/\lambda_{dB})},
\eeq
where $\overline{\lambda}_P$ is the Planck circumference ($=2\pi\lambda_P$), 
with $
\lambda_P = \sqrt{\hbar G /c^3}$, as the Planck length.
The $\lambda$ reduces to $\lambda_{dB}$ for
the low energy regime, and saturates to $4 \lambda_P$ in the 
Planck realm. In this way the Planck scale is not merely 
a dimensional parameter but has been brought in relation
to a universal saturation of
gravitationally-modified de Broglie wavelengths. 

This is a very welcome situation for theories of
quantum gravity where for a long time a paradoxical
situation had existed [\refcite{Magueijo:2001cr,Kowalski-Glikman:2002we}]. 
Each inertial observer
could measure in his frame the fundamental universal 
constants, $\hbar,c,G$, and obtain from  
them a universal  fundamental constant, $\lambda_P$. 
And yet this very $\lambda_P$ \textendash~ being a length scale
\textendash~
is subject to special-relativistic length contraction
which  paradoxically makes it loose its universal character.

The indicated saturation then not only resolves this paradoxical
situation but also suggests that special relativity must suffer
a modification. This modification
must be endowed with the property that it carries two invariant
scales; one the usual $c$, and the second $\lambda_P$. 

The necessity for a  \texttt{SR2} 
as argued in Refs.  [\refcite{Magueijo:2001cr,Kowalski-Glikman:2002we}] is
similar to ours, 
while motivation of Ref.  [\refcite{Amelino-Camelia:2000mn}] is contained in
certain anomalies in astrophysical data; see Refs. 
[\refcite{Ellis:2000sf,Amelino-Camelia:2001qf,Urrutia:2002tr,Sarkar:2002mg,Aloisio:2002ed}].\footnote{Instead 
of the term ``doubly special relativity (DSR)''
coined in the work of Amelino-Camelia,\cite{Amelino-Camelia:2000mn}
we prefer to use the phrase ``special relativity with two 
invariant scales (\texttt{SR2}).'' Without in any way 
questioning  physics content of Amelino-Camelia's proposal,
we take this non-semantic issue  
for the following reason.   The special 
of ``special relativity'' refers to the circumstance that 
one restricts to a special class of inertial observers which move
with a relative uniform velocity. The general of ``general relativity''  
lifts this restrictions. The ``special'' of special relativity
has nothing to do with one versus two
invariants scales. It rather refers to the special class of inertial 
observers; a circumstance that remains unchanged in  special relativity
with two invariant scales. The theory of general relativity with
two invariant scales would thus not be called ``doubly general relativity.''}

\section{Existing \texttt{SR2} proposals}

Simplest of \texttt{SR2} theories result from
keeping the algebra of boost- and 
rotation- generators
intact while modifying the boost parameter in a non-linear manner. 
Specifically, in the \texttt{SR2} of Amelino-Camelia
the boost parameter, $\bv$, changes from the special relativistic form
\beq
\cosh{\varphi} = \frac{E}{m }\,,\quad
\sinh{\varphi} = \frac{p}{m}\,,\quad
\hbv=\frac{\p}{p}\,. \label{dirac}
\eeq
to a new structure\cite{Bruno:2001mw,Judes:2002bw}
\beq
\cosh\xi &=& 
\frac{1}{\mu}\left( \frac{
e^{\lambda_P  E} 
-\cosh\left(\lambda_P\, m \right)}
{\lambda_P \cosh\left( \lambda_P \,m/2\right)}
\right)\,,\label{gac1}\\
\sinh\xi &=& 
\frac{1}{\mu}\left(
\frac{p \,
e^{\lambda_P  E}
}
{\cosh\left(\lambda_P \,m/2\right)} \right) \,,
\quad
\hbxi=\frac{\p}{p}\,, \label{gac2}
\eeq
while for the \texttt{SR2}  of  Magueijo and Smolin the change takes
the form\cite{Magueijo:2001cr,Judes:2002bw}
\beq
\cosh\xi &=& 
\frac{1}{\mu}
\left(\frac{E}{1-\lambda_P\,E}\right)
\,,\label{ms1}\\
\sinh\xi &=& 
\frac{1}{\mu}
\left(\frac{p}{1-\lambda_P\,E}\right)\,,
\quad
\hbxi=\frac{\p}{p}
\,. \label{ms2}
\eeq
Here,  $\mu$ is a Casimir invariant of \texttt{SR2} (see Eq. 
(\ref{ci}) below) and is given by
\beq
\mu =
\cases{ 
\frac{2}{\lambda_P}\,\sinh\left(\frac{\lambda_P \,m}{2}\right) 
& \mbox{for  Ref. [\refcite{Amelino-Camelia:2000mn}]'s \texttt{SR2} }\cr
\frac{m}{1-\lambda_P m}
&  \mbox{for  Ref. [\refcite{Magueijo:2001cr}]'s \texttt{SR2} }}
\eeq
The notation is that of Ref. [\refcite{Judes:2002bw}]; with the minor  exceptions:
$\lambda$, $\mu_0$, $m_0$ there are $\lambda_P$,  $\mu$, $m$ here. 
In what follows we shall 
\textit{generically} represent boost parameter associated
with special relativities with one, or two, invariant scales by $\bx$.
The former relativity shall be abbreviated as \texttt{SR1} 
(to distinguish it from
\texttt{SR2}).\footnote{In this notation the Galilean relativity is 
denoted by \texttt{SR0}.}
Note that giving the explicit expressions for both the 
$\sinh\xi$ and $\cosh\xi$ in Eqs.~(\ref{gac1},\ref{gac2})
is necessary in order to fix the form of the energy-momentum dispersion 
relation through the identity: $\cosh^2\xi-\sinh^2\xi=1$. 
Of course, one may have chosen to work
in terms of one of the hyperbolic trigonometric functions 
\textit{and} the dispersion relation, instead.

At this early stage it is not clear if
there is a unique \texttt{SR2}, or, if 
the final choice will be eventually settled by observational data,
or by some yet-unknown physical principle. Given this ambiguity, this
\textit{Article} addresses itself to presenting a Master equations 
for fermionic and bosonic representations for generic \texttt{SR2}.

\section{Master equation for spin-1/2: Dirac case}
 
Since the underlying 
spacetime symmetry generators remain unchanged much of the  
formal apparatus of the finite dimensional representation
spaces associated with the Lorentz group remains intact.
In particular, there still exist $(1/2,\,0)$ and $(0,\,1/2)$
spinors. But now they transform from the rest frame, to
an inertial frame in which the particle has momentum, $\p$ as:
\beq
\phi_{(1/2,\,0)}\left(\p\right)
& = & \exp\left( + \frac{\bs}{2}\cdot\bx \right)
\phi_{(1/2,0)}\left(\0\right)\,\label{a}\\
\phi_{(0,\,1/2)}\left(\p\right)
& = & \exp\left(- \frac{\bs}{2}\cdot\bx \right)
\phi_{(1/2,0)}\left(\0\right)\,.\label{b}
\eeq  
Since in this \textit{Article}
we do not undertake a study of the behavior of these spinors under
the parity operation, or examine the  massless limit in detail, 
we do not identify the 
$(0,\,1/2)$ spinors as \textit{left-handed} and the $(1/2,\,0)$ spinors
as \textit{right-handed.} Since 
the null momentum vector $\0$ is still isotropic,
one may assume that (see p. 44  of 
Ref. [\refcite{Ryder}] \textit{and} Refs. 
[\refcite{dva_review,Gaioli:1998ra}]): 
\beq
\phi_{(0,1/2)}\left(\0\right) = \zeta \,\phi_{(1/2,0)}\left(\0\right)\,,
\label{c}
\eeq
where $\zeta$ is an undetermined phase factor. 
The analysis presented in Ref. [\refcite{Kirchbach:2002nu}]
 also  convinces
us that the validity of the identity (\ref{c}) is independent of the 
``right-left'' identification of the standard argument.\cite{Ryder,dva_review,Gaioli:1998ra}
In general, the phase
$\zeta$ encodes
C, P, and T properties. The interplay of
Eqs. (\ref{a}-\ref{b}) and (\ref{c}) yields the Master equation for
the $(1/2,\,0)\oplus(0,\,1/2)$ spinors,
\beq
\psi\left(\p\right) = \left(
			\begin{array}{c}
			\phi_{(1/2,\,0)}\left(\p\right)\\
			\phi_{(0,\,1/2)}\left(\p\right)
			\end{array}
		     \right)\,,
\eeq
to be
\beq
\left(
\begin{array}{cc}
-\zeta & \exp\left(\bs\cdot\bx\right) \\
\exp\left(- \bs\cdot\bx\right) & - \zeta^{-1}
\end{array}
\right) \psi\left(\p\right) = 0\,.\label{meq}
\eeq
This is one of the central results of this \textit{Article}.
As a check, taking $\bx$ to be $\bv$, and after some
simple algebraic manipulations,
the Master equation (\ref{meq}) reduces to:
\beq
\left(
 	\begin{array}{cc}
	- m \zeta & E \openone_2 + \bs\cdot \p \\
	E \openone_2 - \bs\cdot \p & - m \zeta^{-1}
	\end{array} 
\right) \,
\psi\left(\p\right) = 0\,,\label{d}
\eeq
where $\openone_n$ stands for $n\times n$ identity matrix
(and $0_n$ shall represent the corresponding null matrix) .
With the given identification of the boost parameter
we are in the realm
of \texttt{SR1}. There, the operation of
parity is well understood. Demanding  parity
covariance for Eq. (\ref{d}), we obtain
$\zeta=\pm 1$. Identifying 
\beq
\left(  \begin{array}{cc}
	0_2 & \openone_2 \\
	\openone_2 & 0_2
	\end{array}
\right)\,,\quad
\left(  \begin{array}{cc}
	0_2 & -\bs\\
	\bs & 0_2
	\end{array}
\right)\,,
\eeq
with the Weyl-representation $\gamma^0$, and $\gamma^i$, respectively,
Eq. (\ref{d}) reduces to the Dirac equation of  \texttt{SR1} 
\beq
\left(\gamma^\mu p_\mu \mp m\right)\psi\left(\p\right)=0\,.\label{de}
\eeq
The linearity of the Dirac equation in, $p_\mu= (E,-\p)$, is now clearly
seen to be associated with two observations: 

\begin{enumerate}
\item[$\mathcal{O}_1$.]
That, $\bs^2 = \openone_2$; and
\item[$\mathcal{O}_2$.]
That in \texttt{SR1}, the hyperbolic functions
\textendash~ 
see Eq. (\ref{dirac}) \textendash~ associated with the boost parameter
are linear in $p_\mu$. 
\end{enumerate}
In \texttt{SR2}, observation $\mathcal{O}_1$
still holds. But, as Eqs. (\ref{gac1} - \ref{gac2}) 
show, $\mathcal{O}_2$ is strongly violated. For this reason the Master equation
(\ref{meq}) cannot be cast in a manifestly covariant form with
a finite number of contracted Lorentz indices of \texttt{SR2} 
as long as we mark spacetime events by $x^\mu$ of \texttt{SR1}.

The last inference is also a welcome result as it indicates
a possible intrinsic non-locality in \texttt{SR2}s. Since in all 
\texttt{SR2}s
the shortest spatial length scales that can be probed are bound
from below by $\lambda_P$, the naively-expected $\delta^3\left(\x-
\x^\prime\right)$ in the anticommutators of the form 
$\left\{\Psi_i\left(\x,t\right),\,\Psi^\dagger_j\left(\x^\prime, t
\right)\right\}$ should be replaced by an highly, but
not infinitely, 
peaked Gaussian-like functions with half-width of the order of
$\lambda_P$.

Note that, the spinors are obtained without reference to a 
wave equation:\footnote{Even though they may also be obtained as a
solution of the relevant wave equation.}

\beq
\psi\left(\p\right) = 
\left(
\begin{array}{cc}
\exp\left( + \frac{\bs}{2}\cdot\bx \right) & \0_2 \\
\0_2 & \exp\left( - \frac{\bs}{2}\cdot\bx \right)
\end{array}
\right)
\left(
			\begin{array}{c}
			\phi_{(1/2,\,0)}\left(\0\right)\\
			\phi_{(0,\,1/2)}\left(\0\right)
			\end{array}
		     \right)\,.
\eeq
The $\phi_{(1/2,\,0)}\left(\0\right)$ 
as well as $\phi_{(0,\,1/2)}\left(\0\right)$ are taken as eigenstates
of the helicity operator,
\beq
\frac{\bs}{2}\cdot\hbxi\,.
\eeq
The choice $\zeta=+1$, in Eq. (\ref{c}), yields the ``particle''
spinors, while,  $\zeta=-1$, gives the ``antiparticle''
spinors.
The extension of the presented formalism for Majorana spinors 
is more subtle.\cite{Majorana:vz,Klapdor-Kleingrothaus:2001ke,Ahluwalia-Khalilova:2003jt,Klapdor-Kleingrothaus:2003gs} 
We hope to present it an extended version of this 
\textit{Article}.

\section{Master equation for higher spins} 

The above-outlined procedure applies
to all, bosonic as well as fermionic,  
$(j,0)\oplus(0,j)$ representation spaces. It is 
not confined to $j=1/2$. A straightforward generalization 
of the $j=1/2$ analysis immediately yields the Master equation 
for an arbitrary-spin,
\beq
\left(
\begin{array}{cc}
-\zeta & \exp\left(2\J\cdot\bx\right) \\
\exp\left(- 2\J\cdot\bx\right) & - \zeta^{-1}
\end{array}
\right) \psi\left(\p\right) = 0\,,\label{j}
\eeq
where
\beq
\psi\left(\p\right) = \left(
			\begin{array}{c}
			\phi_{(j,\,0)}\left(\p\right)\\
			\phi_{(0,\,j)}\left(\p\right)
			\end{array}
		     \right)\,.\label{js}
\eeq
Equation (\ref{j}) 
contains the central result of the previous section as a 
special case.  
For studying the \texttt{SR1} limit it is convenient to bifurcate the
$(j,0)\oplus(0,j)$ space into two sectors by splitting the 
$2(2j+1)$ phases, $\zeta$,
into two sets: $(2j+1)$ phases $\zeta_+$, and the other  
$(2j+1)$ phases $\zeta_-$. Then,  in particle's rest frame
the $\psi(\p)$ may be written as:
\begin{equation}
\psi_h(\0)=
\cases{
u_h(\0)   &  \mbox{when}~$\zeta=\zeta_+$\cr
v_h(\0)   &  \mbox{when}~$ \zeta=\zeta_-$}
\end{equation}
The explicit forms of $u_h(\0)$ and $u_h(\0)$ which we 
shall use (see Eq. (\ref{c})) are:
\beq
u_h(0)=\left(
\begin{array}{c}
\phi_h(\0) \\
\zeta_+ \,\phi_h(\0)   
\end{array}
\right),\,
v_h(0)=
\left(
\begin{array}{c}
\phi_h(\0) \\
\zeta_-\, \phi_h(\0)   
\end{array}
\right),
\eeq
where the   $\phi_h(\0)$ are defined as:
$\J\cdot \hp\, \phi_h(\0) = h \,\phi_h(\0)$, and $h=-j,-j+1,\ldots,+j$.
In the parity covariant \texttt{SR1} 
limit, we find $\zeta_+ = +1$ while   
$\zeta_- = -1$.

As a check, for $j=1$, identification of $\bx$ with $\bv$,
and after implementing parity covariance, yields
\beq
\left(\gamma^{\mu\nu}p_\mu p_\nu \mp m^2\right)\psi(\p)=0\,.\label{bwweq}
\eeq 
The  $\gamma^{\mu\nu}$ are unitarily equivalent  
to those of Ref. [\refcite{Ahluwalia:zt}], and thus we reproduce 
\textit{bosonic matter fields}
with $\left\{C,\,P\right\} = 0$.  A carefully taken massless limit then shows
that the resulting equation is consistent with the free Maxwell equations
of electrodynamics.

Since the $j=1/2$ and $j=1$ representation spaces of \texttt{SR2} reduce to
the Dirac and Maxwell descriptions, it is apparent, that the \texttt{SR2}
contains physics beyond the linear-group realizations of \texttt{SR1}.
To the lowest order in $\lambda_P$,  Eq. (\ref{meq}) yields
\beq
\left(
\gamma^\mu {p}_\mu + \tilde{m} + 
\delta_1\, \lambda_P\right)
\psi(\p)=0\,,
\eeq
where
\beq
\tilde{m}
&=& \left(
\begin{array}{cc}
-\zeta  & 0_2  \\
0_2 & -\zeta^{-1}
\end{array}
\right)\,m \,
\eeq
and
\beq
\delta_1 =
\cases{
\gamma^0\left(\frac{E^2-m^2}{2}\right)+\gamma^i p_i\, E
& \mbox{for  Ref. [\refcite{Amelino-Camelia:2000mn}]'s \texttt{SR2}}\cr
\gamma^\mu p_\mu\, \left(E-m\right)
&       \mbox{for Ref. [\refcite{Magueijo:2001cr}]'s \texttt{SR2}}}
\eeq
Similarly, the presented Master equation can be used  
to obtain \texttt{SR2}'s counterparts for Maxwell's electrodynamic.
Unlike the Coleman-Glashow framework [\refcite{Coleman:1998ti}], 
the principle
of special relativity with two invariant scales provides \textit{all} 
corrections, say, to the standard model of the high energy physics, 
in terms of \textit{one} \textendash~ and 
\textit{not forty six} \textendash~ 
fundamental constant, $\lambda_P$.

\section{Spin-1/2 and Spin-1 description in Judes-Visser 
Variables}
We now take the tentative position,
that the ordinary energy-momentum $p^\mu$ is not the natural physical
variable in \texttt{SR2}s. The Judes-Visser 
variables [\refcite{Judes:2002bw}]: $\eta^\mu \equiv 
\left(\epsilon(E,p),\,\bp(E,p)\right)=(\eta^0,\be)$
appear more suited to describe physics sensitive to Planck scale.
The $\epsilon(E,p)$ and $\bp(E,p)$ 
relate to the rapidity parameter $\bx$ of  \texttt{SR2} 
in same functional form
as do $E$ and $\p$ to  $\bv$ of \texttt{SR1}:
\beq
\cosh\left(\xi\right)= \frac{\epsilon(E,p)}{\mu}\,,\quad
\sinh\left(\xi\right)= \frac{\pi(E,p)}{\mu}\,,
\eeq
where 
\beq
\mu^2= \left[\epsilon(E,p)\right]^2 - \left[\bp(E,p)\right]^2\,.
\label{ci}
\eeq
They provide 
the most economical and physically transparent formalism for 
representation space theory in \texttt{SR2}.   
For $j=1/2$ and $j=1$, Eq. (\ref{j}) yields the \textit{exact}
SR2 equations for $\psi(\bp)$:
\beq
&&\left(\gamma^\mu \eta_\mu + \tilde{\mu}\right)
\psi\left(\bp\right)=0\,,\label{denew}\\
&&\left(\gamma^{\mu\nu}\eta_\mu \eta_\nu + \tilde{\mu}^2\right)
\psi(\bp)=0\,,\label{bwweqnew}
\eeq
where
\beq
\tilde{\mu}
&=& \left(
\begin{array}{cc}
-\zeta^{-1}  & 0_2 \\
0_2 & -\zeta
\end{array}
\right)\,\mu\,.
\eeq

\section{Concluding Remarks}

Our task in this \textit{Article} was to provide a description 
of fermions and bosons at the  level of representation space 
theory in \texttt{SR2}. However, we confined entirely to the
representations of the type $(j,0)\oplus(0,j)$ \textendash~ 
these types are important for matter fields, and 
to study gauge-field strength tensors.
To study \texttt{SR2}'s effect on the gauge fields and  weak-field gravity
the present \textit{Article's} formalism needs to be extended to 
$(j,j)$ representation spaces.
In view of Weinberg's earlier works [\refcite{Weinberg:ew}] it is known
that there is a deep connection between \textit{local} quantum field theory,  
\textit{SR1}  $(j,j)$  spaces [\refcite{Kirchbach:2002nu}], and the \textit{equality} 
of the inertial and gravitational masses. Therefore, the suggested study
must answer \texttt{SR2}'s effect on the equivalence principle.

In quantum field theoretic framework, 
the  special relativity's spacetime $x^\mu$ is canonically conjugate
to $p_\mu$, and appears in the  field operators as:
\beq
\Psi(x)= \int \frac{d^3 \p}{(2\pi)^3} \frac{m}{p_0} 
\sum_{h=-j}^{+j}{\Big[}
a_h(\p) u_h(\p) e^{- i p_\mu x^\mu}  +
b_h(\p) v_h(\p)
e^{i p_\mu x^\mu}{\Big]}\,,
\eeq
where the particle-antiparticle spinors, $u_h(\p)$
and $v_h(\p)$ (generically represented by $\psi_h(\p)$), 
are solutions of the Master equations (but with 
$\bx\rightarrow\bv$)
introduced above, and can be readily obtained from:
\beq
\psi_h(\p)=
\left(
\begin{array}{cc}
\exp(+ \J\cdot\bv) & 0_{2j+1} \\
 0_{2j+1} & \exp(-\J\cdot\bv)
\end{array}
\right)\psi_h(\0)\,.
\eeq
Now, as our discussion on non-locality indicates 
$x^\mu$ of  \texttt{SR1} is perhaps not the natural 
physical spacetime variable
at the Planck scale. The spacetime at Planck scale, we suggest,
is represented by new event vectors $\chi^\mu$ (to be treated as
``canonically conjugate'' to Judes-Visser variable $\eta_\mu$); and
suggests the following definition for the field operators
built upon the \texttt{SR2}'s spinors:
\beq
\Psi(\chi)= \int \frac{d^3 \be}{(2\pi)^3} \frac{\mu}{\eta_0} 
\sum_{h=-j}^{+j}{\Big[}
a_h(\be) u_h(\be)
e^{- i \eta_\mu \chi^\mu}   +
b_h(\be) v_h(\be)
e^{i \eta_\mu \chi^\mu}{\Big]}\,,
\eeq
with 
\beq
\psi_h(\be)=
\left(
\begin{array}{cc}
\exp(+ \J\cdot\bx) & 0_{2j+1} \\
 0_{2j+1} & \exp(-\J\cdot\bx)
\end{array}
\right)\psi_h(\0)\,.
\eeq
Immediately, we verify that for spin-$1/2$ fermions in \texttt{SR2}
\beq
\left\{\Psi_i\left(\bc,\chi^0\right),\,\Psi^\dagger_j\left(\bc^\prime,
\chi^0\right)\right\} 
=\delta^3
\left(\bc-\bc^\prime\right)\,\delta_{ij}\,.
\eeq
What appears as non-locality in the space of events marked 
by $x^\mu$ now,  in the space of events marked by 
$\chi^\mu$,  exhibits itself as
locality.  This is a rather unexpected observation and it
calls for a deeper understanding of 
the $\eta_\mu$ and $\chi^\mu$ description of \texttt{SR2}. 
The Planck length is intrinsically built in the
latter spacetime variables, and it may carry significant relevance for
extending \texttt{SR2} to the gravitational realm. 

The evolution of special relativity in the sequence\footnote{The symbols 
above the arrows indicate the invariants for the subsequent 
\texttt{SRn}.}  
\beq
\mbox{\texttt{SR0}}
\stackrel{c}{\longrightarrow}
\mbox{\texttt{SR1}}
\stackrel{c,\lambda_P}{\longrightarrow}
\mbox{\texttt{SR2}}
\eeq
 translates to giving spacetime, first, a 
\textit{relativistic} and, then, a \textit{quantum-gravitational} 
character. The work initiated here, and in Ref. [\refcite{Arzano:2002yd}], gives concrete
shape to modifications that one may expect in the standard model
of high-energy physics and theory of gravitation.

\section{Addenda: A brief Critique}

This \textit{Article} was penned sometime ago and requires some 
remarks in the form of an addenda and a critique.
These are enumerated below:

\begin{enumerate}

\item
If all turns out as claimed in this \textit{Article}
then \texttt{SR2} shall constitute a fundamentally new program 
for  a theory of quantum gravity. At present, there exist serious 
questions on physical distinguishability of \texttt{SR2}'s
from SR1. I have already written on the subject
elsewhere [\refcite{Ahluwalia-Khalilova:2002wf}]
and a number of other authors have raised similar questions, see, e.g.,
Refs. 
[\refcite{Grumiller:2003df,Schutzhold:2003yp,Toller:2003tz,Jafari:2003xt}].
To brush aside these issues with an argument, such as, 
``Mathematical triviality by no means implies physical equivalence, 
and one may argue that it is in fact an asset,''\cite{Magueijo:2003gj} 
only delays 
resolution of the issues involved.
As has been noted by Czerhoniak\cite{Czerhoniak:2003nb} the 
question of distinguishability of  \texttt{SR2} and  \texttt{SR1} 
is deeply connected with observations of Lukierski and Nowicki
[\refcite{Lukierski:2002df,Lukierski:2002wf}] \textemdash~i.e., 
whether or not
the underlying momentum-space/spacetime is commutative, 
or non-commutative. I have argued in Ref. [\refcite{Ahluwalia:dd}]
that the latter is not a matter of choice but a logical 
implication of the interface of the gravitational and quantum
frameworks. The question then is what precise 
form this non-commutativity takes and if this too, in some way, can
also be deciphered from a critical study of representation
spaces associated with the spacetime symmetries. 
A tentative answer, to be presented
elsewhere, is in the affirmative. Surprisingly, the non-commutativity seems
to depend on the spin content and C, P, and T properties of the
examined representation space.  Therefore,
in a physically successful \texttt{SR2},
the notion of spacetime is expected to be deeply 
intertwined with specific properties of the test particle.

\item
In Ref. [\refcite{Arzano:2002yd}], the authors have brushed aside 
the importance of a relative phase difference \textemdash~
specifically in the notation of Eq. (\ref{c}), they ignore $\zeta = -1$.
This is of more than an academic interest. Without it, all antiparticle
are absent from the theory.\footnote{Unless one ignores a 
set of two mistakes which
\textit{partly} cancel out the effect, but it then comes back to plague
when one considers Maxewell's field, or gravitation.\cite{Ahluwalia:zt}}

\end{enumerate}

\section*{Acknowledgments}

I thank Giovanni Amelino-Camelia, 
Gaetano Lambiase, and their respective institutes
in Rome and Salerno, for discussions and hospitality
in May 2001. I  thank Daniel Grumiller for a series of discussions 
on this subject.

\textit{
This work is supported by Consejo Nacional de Ciencia y
Tecnolog\'ia (CONACyT, Mexico).}


\begin{thebibliography}{999}


\bibitem{Kempf:1994su}
A.~Kempf, G.~Mangano and R.~B.~Mann,
``Hilbert space representation of the minimal length uncertainty relation,''
Phys.\ Rev.\ D {\bf 52} (1995) 1108
[arXiv:hep-th/9412167].


\bibitem{Ahluwalia:2000iw}
D.~V.~Ahluwalia,
 ``Wave-Particle duality at the Planck scale: Freezing of neutrino oscillations,''
Phys.\ Lett .\ A {\bf 275} (2000) 31
[arXiv:gr-qc/0002005].


\bibitem{Magueijo:2001cr}
J.~Magueijo and L.~Smolin,
``Lorentz invariance with an invariant energy scale,''
Phys.\ Rev.\ Lett.\  {\bf 88} (2002) 190403
[arXiv:hep-th/0112090].


\bibitem{Kowalski-Glikman:2002we}
J.~Kowalski-Glikman and S.~Nowak,
 ``Doubly special relativity theories as different bases of kappa-Poincare
algebra,''
Phys.\ Lett.\ B {\bf 539} (2002) 126
[arXiv:hep-th/0203040].


\bibitem{Amelino-Camelia:2000mn}
G.~Amelino-Camelia,
 ``Relativity in space-times with short-distance structure governed by an
observer-independent (Planckian) length scale,''
Int.\ J.\ Mod.\ Phys.\ D {\bf 11} (2002) 35
[arXiv:gr-qc/0012051].


\bibitem{Ellis:2000sf}
J.~R.~Ellis, N.~E.~Mavromatos and D.~V.~Nanopoulos,
``Space-time foam effects on particle interactions and the GZK cutoff,''
Phys.\ Rev.\ D {\bf 63} (2001) 124025
[arXiv:hep-th/0012216].



\bibitem{Amelino-Camelia:2001qf}
G.~Amelino-Camelia,
``Space-time quantum solves three experimental paradoxes,''
Phys.\ Lett.\ B {\bf 528} (2002) 181
[arXiv:gr-qc/0107086].

\bibitem{Urrutia:2002tr}
L.~Urrutia,
``Quantum gravity corrections to particle interactions,''
Mod.\ Phys.\ Lett.\ A {\bf 17} (2002) 943
[arXiv:gr-qc/0205103].


\bibitem{Sarkar:2002mg}
S.~Sarkar,
``Possible astrophysical probes of quantum gravity,''
Mod.\ Phys.\ Lett.\ A {\bf 17} (2002) 1025
[arXiv:gr-qc/0204092].

\bibitem{Aloisio:2002ed}
R.~Aloisio, P.~Blasi, A.~Galante, P.~L.~Ghia and A.~F.~Grillo,
``Space time fluctuations and ultra high energy cosmic ray interactions,''
Astropart.\ Phys.\  {\bf 19} (2003) 127
[arXiv:astro-ph/0205271].



\bibitem{Bruno:2001mw}
N.~R.~Bruno, G.~Amelino-Camelia and J.~Kowalski-Glikman,
``Deformed boost transformations that saturate at the Planck scale,''
Phys.\ Lett.\ B {\bf 522} (2001) 133
[arXiv:hep-th/0107039].


\bibitem{Judes:2002bw}
S.~Judes and M.~Visser,
``Conservation Laws in Doubly Special Relativity,''
Phys.\ Rev.\ D {\bf 68} (2003) 045001
[arXiv:gr-qc/0205067].



\bibitem{Ryder}
L. H. Ryder, \textit{Quantum Field Theory} (Cambridge University Press,
Cambridge, 1987). The reader looking at the Second Edition (1996) should
refer to p. 41, instead of p.44.

\bibitem{dva_review}
D. V. Ahluwalia, Found. Phys. 
{\bf 28}, 527 (1998);
D. V. Ahluwalia, M. Kirchbach, 
Int. J. Mod. Phys. D \textbf{10}, 811 (2001). 


\bibitem{Gaioli:1998ra}
F.~H.~Gaioli and E.~T.~Garcia Alvarez,
 ``Some remarks about intrinsic parity in Ryder's derivation of the Dirac
equation,''
Am.\ J.\ Phys.\  {\bf 63} (1995) 177
[arXiv:hep-th/9807211].


\bibitem{Kirchbach:2002nu}
M.~Kirchbach and D.~V.~Ahluwalia,
``Spacetime structure of massive gravitino,''
Phys.\ Lett.\ B {\bf 529} (2002) 124
[arXiv:hep-th/0202164];
D.~V.~Ahluwalia and M.~Kirchbach,
``(1/2,1/2) representation space: An ab initio construct,''
Mod.\ Phys.\ Lett.\ A {\bf 16} (2001) 1377
[arXiv:hep-th/0101009].


\bibitem{Majorana:vz}
E.~Majorana,
``Theory Of The Symmetry Of Electrons And Positrons,''
Nuovo Cim.\  {\bf 14} (1937) 171.



\bibitem{Klapdor-Kleingrothaus:2001ke}
H.~V.~Klapdor-Kleingrothaus, A.~Dietz, H.~L.~Harney and I.~V.~Krivosheina,
Mod.\ Phys.\ Lett.\ A {\bf 16} (2001) 2409
[arXiv:hep-ph/0201231].




\bibitem{Ahluwalia-Khalilova:2003jt}
D.~V.~Ahluwalia-Khalilova,
 ``Extended set of Majorana spinors, a new dispersion relation, and a
preferred frame,''
arXiv:hep-ph/0305336.






\bibitem{Klapdor-Kleingrothaus:2003gs}
H.~V.~Klapdor-Kleingrothaus, A.~Dietz, I.~V.~Krivosheina, C.~Doerr and C.~Tomei,
``Support of evidence for neutrinoless double beta decay,''
Phys.\ Lett.\ B {\bf 578} (2004) 54
[arXiv:hep-ph/0312171].


\bibitem{Ahluwalia:zt}
D.~V.~Ahluwalia, M.~B.~Johnson and T.~Goldman,
``A Bargmann-Wightman-Wigner Type Quantum Field Theory,''
Phys.\ Lett.\ B {\bf 316} (1993) 102
[arXiv:hep-ph/9304243].


\bibitem{Coleman:1998ti}
S.~R.~Coleman and S.~L.~Glashow,
``High-energy tests of Lorentz invariance,''
Phys.\ Rev.\ D {\bf 59} (1999) 116008
[arXiv:hep-ph/9812418].


\bibitem{Weinberg:ew}
S.~Weinberg,
 ``Photons And Gravitons In S Matrix Theory: Derivation Of Charge Conservation
And Equality Of Gravitational And Inertial Mass,''
Phys.\ Rev.\  {\bf 135} (1964) B1049.

\bibitem{Arzano:2002yd}
M.~Arzano and G.~Amelino-Camelia,
``Dirac spinors for doubly special relativity,''
arXiv:gr-qc/0207003.

\bibitem{Ahluwalia-Khalilova:2002wf}
D.~V.~Ahluwalia-Khalilova,
 ``Operational indistinguishabilty of doubly special relativities from  special
arXiv:gr-qc/0212128.




\bibitem{Grumiller:2003df}
D.~Grumiller, W.~Kummer and D.~V.~Vassilevich,
``A note on the triviality of kappa-deformations of gravity,''
Ukr.\ J.\ Phys.\  {\bf 48} (2003) 329
[arXiv:hep-th/0301061].



\bibitem{Schutzhold:2003yp}
R.~Schutzhold and W.~G.~Unruh,
``Problems of doubly special relativity with variable speed of light,''
arXiv:gr-qc/0308049.
See also,
M.~Arzano,
``Comment on 'large-scale non-locality in doubly special relativity with an
energy-dependent speed of light',''
arXiv:gr-qc/0309077.


\bibitem{Toller:2003tz}
M.~Toller,
``On the Lorentz transformations of momentum and energy,''
Mod.\ Phys.\ Lett.\ A {\bf 18} (2003) 2019
[arXiv:hep-th/0301153].

\bibitem{Jafari:2003xt}
N.~Jafari and A.~Shariati,
``Comments on varying speed of light theories,''
arXiv:gr-qc/0312007.



\bibitem{Magueijo:2003gj}
J.~Magueijo,
``New varying speed of light theories,''
Rept.\ Prog.\ Phys.\  {\bf 66} (2003) 2025
[arXiv:astro-ph/0305457].


\bibitem{Czerhoniak:2003nb}
P.~Czerhoniak,
``Twisting kappa-deformed phase space,''
J.\ Phys.\ A {\bf 36} (2003) 9655
[arXiv:hep-th/0302036].


\bibitem{Lukierski:2002df}
J.~Lukierski and A.~Nowicki,
 ``Doubly Special Relativity versus $\kappa$-deformation of relativistic
kinematics,''
Int.\ J.\ Mod.\ Phys.\ A {\bf 18} (2003) 7
[arXiv:hep-th/0203065].

\bibitem{Lukierski:2002wf}
J.~Lukierski and A.~Nowicki,
 ``Nonlinear and Quantum Origin of Doubly Infinite Family of Modified Addition
Czech.\ J.\ Phys.\  {\bf 52} (2002) 1261
[arXiv:hep-th/0209017].



\bibitem{Ahluwalia:dd}
D.~V.~Ahluwalia,
``Quantum Measurements, Gravitation, And Locality,''
Phys.\ Lett.\ B {\bf 339} (1994) 301
[arXiv:gr-qc/9308007].



\end{thebibliography}
\end{document}